\documentclass[aip,pof]{revtex4-1}

\usepackage{graphicx}
\usepackage{dcolumn}
\usepackage{bm}
\usepackage{psfrag}

\begin{document}

\preprint{APS/123-QED}

\title{Transition to finger convection in double-diffusive convection}

\author{M. Kellner}

\author{A. Tilgner}

\affiliation{Institute of Geophysics, University of G\"ottingen,
Friedrich-Hund-Platz 1, 37077 G\"ottingen, Germany }

\date{\today}

\begin{abstract}
Finger convection is observed experimentally in an electrodeposition cell in
which a destabilizing gradient of copper ions is maintained against a
stabilizing temperature gradient. This double-diffusive system shows finger
convection even if the total density stratification is unstable. Finger
convection is replaced by an ordinary convection roll if convection is fast
enough to prevent sufficient heat diffusion between neighboring fingers, or if
the thermal buoyancy force is less than 1/30 of the compositional buoyancy
force. At the transition, the ion transport is larger than without an opposing
temperature gradient.
\end{abstract}

\maketitle

\section{Introduction}

If convection occurs in natural systems, the density of the convecting fluid is
frequently affected by several quantities. The most widely studied example is
convection in ocean water, where both temperature and salinity determine the
density. Different flow structures may occur in convecting ocean water as
opposed to distilled water because heat and salinity have different
diffusivities \cite{Stern60,Turner74,Turner85,Schmitt94}. Salt fingers are the best known of
these double-diffusive structures. Fingers are narrow vertical columns in which
fluid is moving vertically. They require a stabilizing temperature and a
destabilizing salt concentration gradient. They allow convective transport even
if the total density stratification is stable. Fingers form because heat
diffuses more rapidly than ions. The fingers must be narrow enough so
that heat diffuses between neighboring fingers, but broad enough so that
salinity cannot be significantly exchanged between one finger and the next\cite{Turner67,Linden73,Schmitt83}.

According to the commonly accepted picture for finger formation, narrow fingers should
not appear if the stabilizing temperature gradient is weak enough so that
density increases with height due to the destabilizing salinity gradient. In
this case, a convection roll of the same form as observed in ordinary
Rayleigh-B\'enard convection is expected. These convection rolls have in order
of magnitude the same width and height and therefore suffer less from
dissipative losses as long and narrow fingers. One naively expects convection
rolls to supercede fingers as long as the fluid is top heavy. It thus came as a
surprise that a recent experiment found finger convection even in unstably
stratified fluids with a weak stabilizing temperature gradient \cite{Hage10}.
The present paper explores more thoroughly the conditions for the existence of
fingers. It is shown that fingers in the top heavy fluid are really a
form of convection distinct from a convection roll and that convection undergoes
a genuine transition as control parameters are varied and the system switches
from fingers to rolls. This removes doubts that fingers are only a metastable
form of convection which disappears if the system is given enough time. The new
measurements reported here locate the transition in parameter space and allow us
to test various hypotheses concerning the necessary conditions for finger
formation.

The next section summarizes the experimental procedures. Apart from the use of
thermochromic liquid crystals and minor improvements in temperature control, the
apparatus employed here is exactly the same as in Ref. \onlinecite{Hage10}. Full
details of the system are given there and will only be repeated in the next
section to the extent necessary to make the paper self contained. The third
section presents the results.

\section{The Experiment}

The experiments maintain thermal and ion concentration gradients across a
convecting fluid by an electrochemical technique \cite{Hage10,Goldst90}. A cell
made of copper top and bottom plates and plexiglass side walls is filled with a
dilute solution of $CuSO_4$ in sulfuric acid. The copper plates serve both as
temperature controlled boundaries and as electrodes. When a potential difference
is applied between the two electrodes, a current flows through the cell with
copper ions dissolving from one electrode and reattaching to the other. The
sulfuric acid does not participate in the electrochemical reaction, but its free
ions screen the electric field from the bulk of the cell. Apart from microscopic
boundary layers, the copper ions diffuse and are advected through the cell, but
do not experience any electric field. The temperature and the copper ion
concentration are the two agents determining the density of the fluid.

The material properties entering the problem of double-diffusive convection are
the kinematic viscosity of the fluid $\nu$, the diffusivities of temperature and
ion concentration, $\kappa$ and $D$, and two expansion coefficients $\alpha$ and
$\beta$ determining variations of density $\rho$ as a function of temperature
$T$ and copper ion concentration $c$ around a reference state with density,
temperature, concentration and pressure $\rho_0$, $T_0$, $c_0$ and $p_0$ via
\begin{equation}
\alpha=-\frac{1}{\rho_0}\left(\frac{\partial \rho}{\partial T}
\right)_{c_0,\rho_0,p_0}
~~~,~~~
\beta=\frac{1}{\rho_0}\left(\frac{\partial \rho}{\partial c}
\right)_{T_0,\rho_0,p_0}.
\end{equation}
Both $\alpha$ and $\beta$ are positive. Additional control parameters are the
gravitational acceleration $g$, the cell height $L$, and the temperature and
concentration differences applied across the cell, $\Delta T$ and $\Delta c$,
defined as 
\begin{equation}
\Delta T = T_{\rm{bottom}}-T_{\rm{top}} ~~~,~~~ \Delta c = c_{\rm{top}}-c_{\rm
{bottom}}
\end{equation}
with the subscripts indicating the boundary at which temperature $T$ or
concentration $c$ are evaluated. The concentration difference $\Delta c$ is only
known if the cell is operated at the limiting current \cite{Hage10,Goldst90}.
In that situation, $\Delta c = 2 c_0$, where $c_0$ is the average concentration
of copper in the solution.

Four non-dimensional numbers are necessary to specify all control parameters of
double-diffusive convection. We will use the Prandtl and Schmidt numbers, $Pr$
and $Sc$, defined as
\begin{equation}
Pr=\frac{\nu}{\kappa} ~~~,~~~ Sc=\frac{\nu}{D}.
\end{equation}
These are material properties which we will consider to be constants for the
purpose of this paper with $Pr \approx 8.7$ and $Sc \approx 1970$. These average
numbers are slightly different from those given in ref. \onlinecite{Hage10}
because the measurements presented here focus on a transition and sample only a
subspace of the accessible parameter space. A summary of the data obtained near
the transition is given in table \ref{table1}. For all the measurements listed in
this table, $Pr$ varies between 8.55 and 8.87, and $Sc$ varies between 1910 and
2050.

The driving
forces are characterized by the thermal and chemical Rayleigh numbers, $Ra_T$
and $Ra_c$, given by
\begin{equation}
Ra_T=\frac{g \alpha \Delta T L^3}{\kappa \nu}  ~~~,~~~ 
Ra_c=\frac{g \beta \Delta c L^3}{D \nu}.
\end{equation}
With the sign conventions introduced above, a negative Rayleigh number indicates
a stable stratification, which means that in the finger regime, $Ra_T$ is
negative and $Ra_c$ positive. Another quantity of interest is the density ratio
$\Lambda$ which quantifies the ratio of thermal and chemical buoyancy and is
given by
\begin{equation}
\Lambda=\frac{Ra_T}{Ra_c} \frac{\kappa}{D} = \frac{\alpha \Delta T}{\beta \Delta
c}.
\end{equation}

Among the observables is the Sherwood number, which is
directly proportional to the number of ions transported from top to bottom
divided by the purely diffusive current, so that the Sherwood number can be
determined by
\begin{equation}
    Sh=\frac{j\,L}{z\,F\,D\,\Delta c}
\end{equation}
if $j$ is the current density, $z$ the valence of the ion ($z=2$ for $Cu^{2+}$)
and $F$ Faraday's constant. The Sherwood number plays the same role for the ion
transport as the Nusselt number for the heat transport. Measurements of the
Nusselt number have not been attempted, as they would require a more accurate
temperature monitoring than was implemented in the apparatus.

The velocity field was characterized by PIV. A vertical plane near the middle of
the cell was illuminated with a pulsed laser and observed at right angles with a
camera. Based on the horizontal and vertical components of velocity in the plane
of illumination, three different Reynolds numbers, $Re_x$, $Re_y$, and $Re$ will
be useful:
\begin{equation}
Re_x=\frac{L}{\nu} \left( \frac{1}{A} \int v_x^2 dA \right)^{1/2}  ~~~,~~~
Re_y=\frac{L}{\nu} \left( \frac{1}{A} \int v_y^2 dA \right)^{1/2}  ~~~,~~~
Re=\sqrt{Re_x^2+Re_y^2}.
\end{equation}
The integrals extend over the surface $A$ pictured during the PIV measurements,
which typically reaches from plate to plate in the vertical and covers more than half
of the cell width in the horizontal. The PIV measurements, together with
shadowgraph pictures, allow the measurement of the finger thickness $d$.

For convenient reference, we repeat here the scaling laws given in Ref.
\onlinecite{Hage10} for finger convection:
\begin{equation}
\frac{d}{L} = 0.95 |Ra_T|^{-1/3} Ra_c^{1/9}
\label{eq_d_Ra}
\end{equation}
\begin{equation}
Re=10^{-6} |Ra_T|^{-1/2} Ra_c
\label{eq_Re_Ra}
\end{equation}
\begin{equation}
Sh=0.016 |Ra_T|^{-1/12} Ra_c^{4/9}
\label{eq_Sh_Ra}
\end{equation}
In the course of the new experiments, some of the measurements reported in table
1 of Ref. \onlinecite{Hage10} were reproduced. However, the entry with $L=2cm$,
$\Delta T/L=-0.1 K/cm$ could not be reproduced,
presumably because of insufficient temperature control in the earlier
experiment. At those parameters, no fingers were detected in the new experiments.

Thermochromic liquid crystals have been used in a few of the new experiments in
order to obtain some information about the temperature field. For those
experiments, encapsulated liquid crystal particles of diameters around $50 \mu
m$ were suspended in the electrolyte and a vertical plane was illuminated
with white light and observed at right angles with a CCD camera. The pictures
taken by the camera were encoded at each pixel in terms of the three variables
hue, saturation and intensity. The hue carries the information about the
temperature \cite{Dabiri91}. The dependence of hue on temperature is conveniently
determined by taking a picture of the cell with the uniform temperature gradient
established before the voltage is applied to the cell. The temperatures of the
plates were adjusted to exploit as well as possible the color play of the liquid
crystal.

\begin{figure}
\includegraphics[width=8cm]{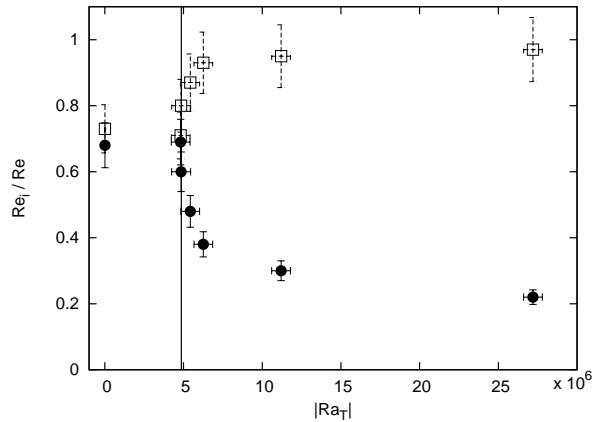}
\caption{$Re_x/Re$ (filled circles) and $Re_y/Re$ (empty squares) as a function of $|Ra_T|$ for
$Ra_c=2.92 \times 10^{10}$. The vertical line marks the transitional $|Ra_T|$
deduced from this figure which reappears as one data point in fig. \ref{fig_Phasenraum}.}
\label{fig_ReivsRa}
\end{figure}

\begin{figure}
\includegraphics[width=7.2cm]{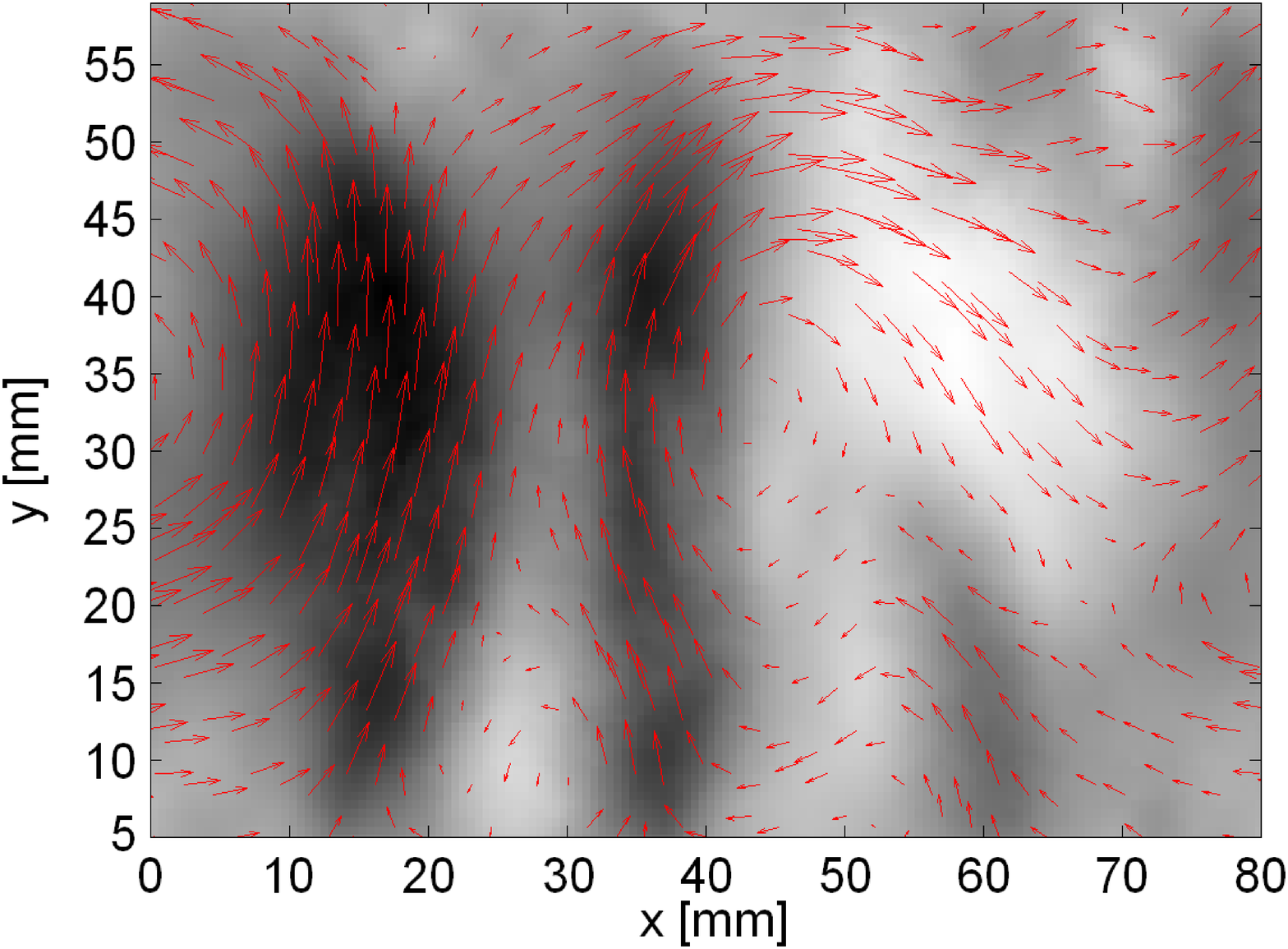}
\includegraphics[width=7.2cm]{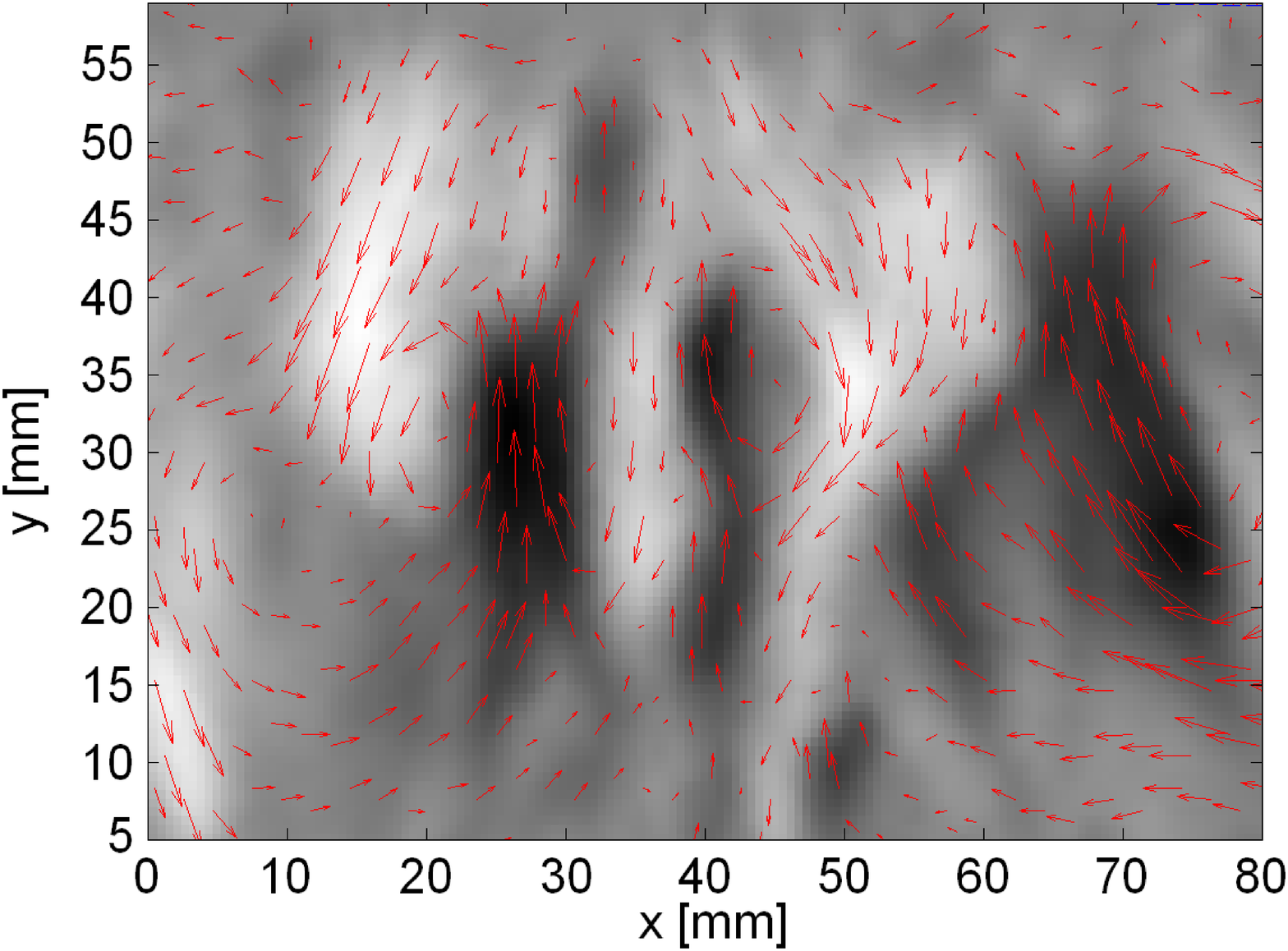}\\
\includegraphics[width=7.2cm]{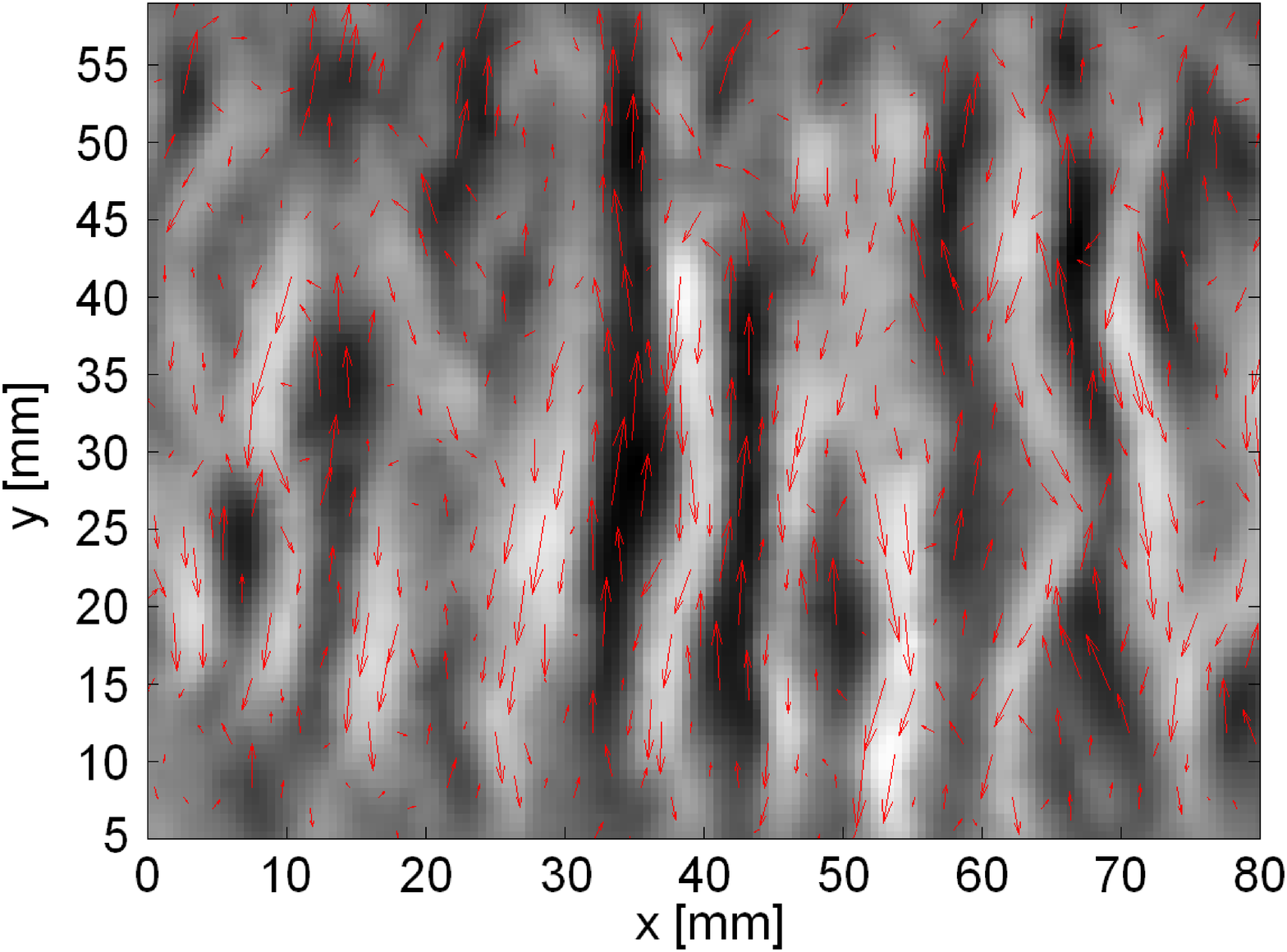}
\includegraphics[width=7.2cm]{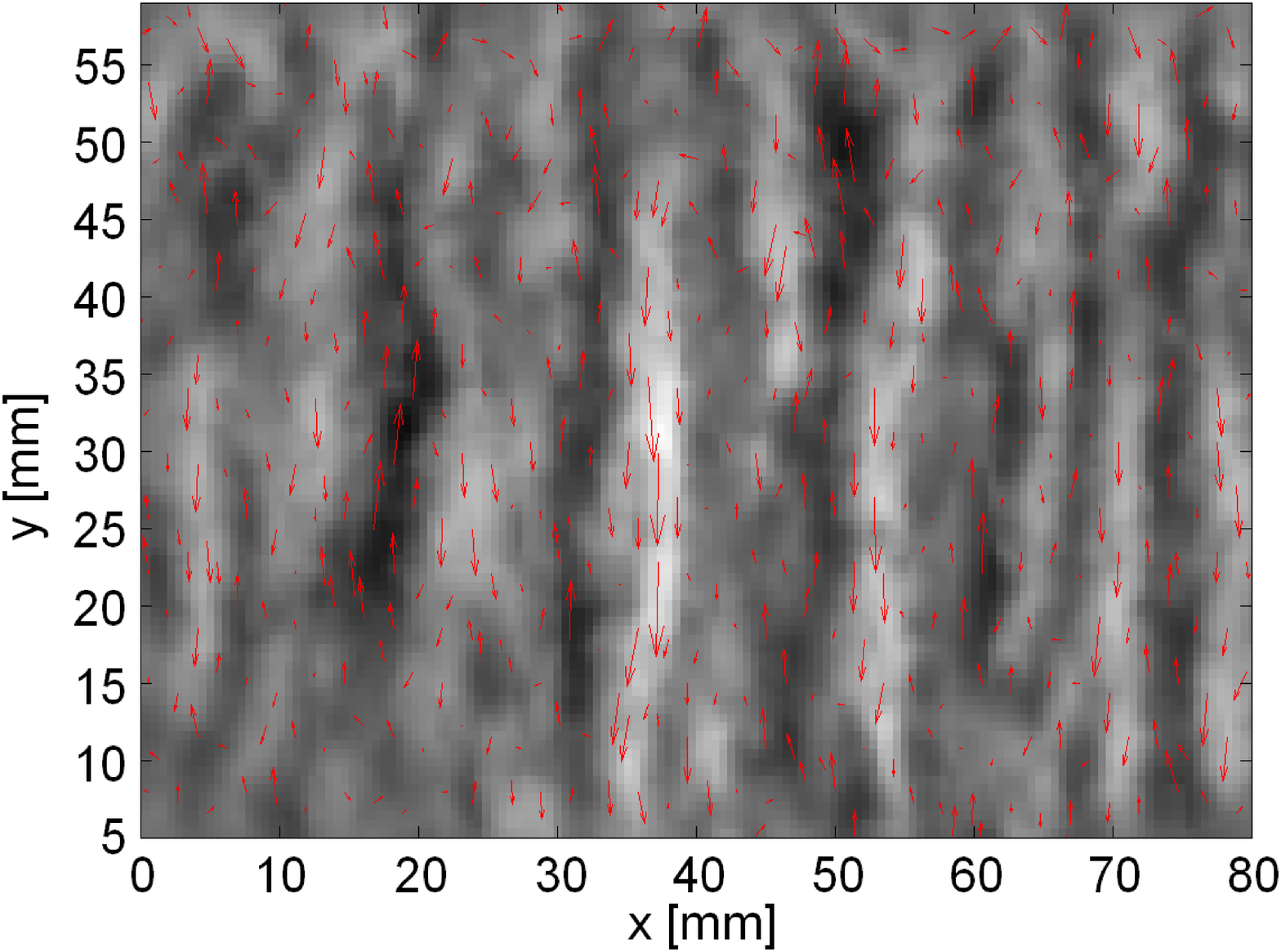}
\caption{(Color online) Transition for $Ra_c=2.92\cdot 10^{10}$. Shown is the 
vertical component of the
velocity field at $|Ra_T|=5.94\cdot10^3$ (top left), $|Ra_T|=4.85\cdot10^6$ (top right), 
$|Ra_T|=1.12\cdot10^7$ (bottom left) and $|Ra_T|=2.72\cdot10^7$ (bottom right). The (red)
arrows indicate the velocity vector, and the grayshade depends on the vertical velocity
component (dark gray in upflows and light gray in downflows). All panels show
the velocity field in a vertical plane near the center of the experimental cell.
The top and bottom plates are located at $y=0\;$mm and $y=60\;$mm, respectively, and the total
lateral extent of the cell is $200 \;$mm.}
\label{fig_vectorfield}
\end{figure}

\begin{figure}
\includegraphics[width=8cm]{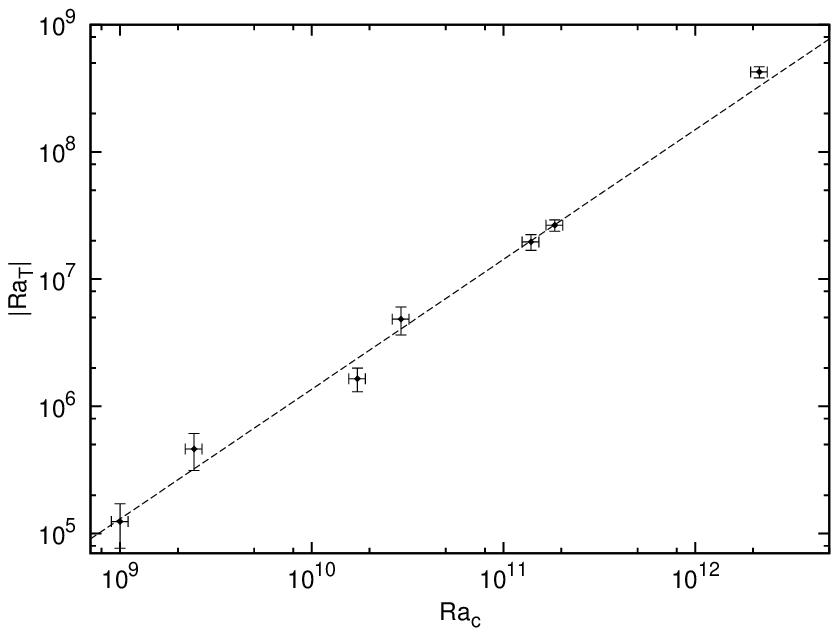}
\caption{The transition line between finger convection and convection rolls in
the $|Ra_T|,Ra_c-$plane. The symbols indicate measurements, and the dotted line
is a fit given by $|Ra_T|=8.6 \times 10^{-5} Ra_c^{1.02}$.}
\label{fig_Phasenraum}
\end{figure}

\section{Results and discussion}

It appears from the scaling of the finger width (\ref{eq_d_Ra}) that fingers
broaden with decreasing stabilizing $|Ra_T|$, and that they possibly
continuously transform into convection rolls, so that there is no transition
between two fluid dynamic states, the fingers and the convection rolls. However,
this is not true and there is a genuine transition from the finger regime to
simple convection. It is convenient to demonstrate the existence of a transition
with the help of the Reynolds numbers $Re_x$ and $Re_y$ based on the horizontal
and vertical velocity components alone. Fig. \ref{fig_ReivsRa} shows $Re_x/Re$
and $Re_y/Re$ as a function of $|Ra_T|$ at fixed $Ra_c$. Fingers exist for large
$|Ra_T|$. The velocity field is then very anisotropic with a vertical 
velocity much larger than the horizontal velocity almost everywhere. For small
$|Ra_T|$, the fingers disappear and are replaced by a convection roll with an
approximately isotropic velocity field with $Re_x/Re \approx Re_y/Re \approx
1/\sqrt{2}$. Fig. \ref{fig_vectorfield} provides a visual impression of the
various flows encountered during the transition.
As fig. \ref{fig_ReivsRa} shows, the transition from one state to
the other occurs at a well defined $|Ra_T|$.

These measurements have been obtained by setting $Ra_T$ to a fixed value and
letting the temperature gradient establish itself by diffusion before the
voltage was applied to the cell. A few experimental runs were made to exclude
the possibility of a hysteresis. In these experiments, the cell was first left to
equilibrate in a convection state at a $|Ra_T|$ below the transition to the
finger state, before $|Ra_T|$ was suddenly increased to a value above the
transition. It was also checked for the opposite direction that no hysteresis
appears. Fingers always reappeared
with the same amplitude and characteristics as before. In summary, the transition
from a convection roll to finger convection is well described as a supercritical
bifurcation.

\begin{figure}
\includegraphics[width=8cm]{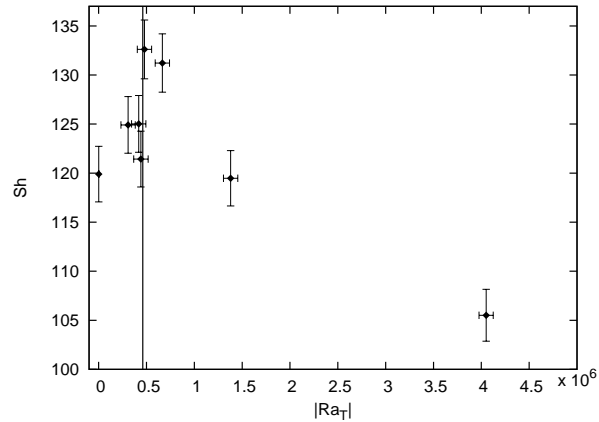}
\caption{$Sh$ as a function of $|Ra_T|$ for $Ra_c=2.43 \times 10^{9}$. The
vertical line marks the transition determined from a measurement as shown in
fig. \ref{fig_ReivsRa}.}
\label{fig_ShvsRa}
\end{figure}

Measurements as shown in fig. \ref{fig_ReivsRa} have been repeated at six more
chemical Rayleigh numbers. The transition line in the $|Ra_T|,Ra_c-$plane is
shown in fig. \ref{fig_Phasenraum}. The measurements are compatible with a
transition line defined by $Ra_T/Ra_c = const.$, or $\Lambda=-1/30$. The best
approximation to the transition line is a power law given by $|Ra_T|=8.6 \times
10^{-5} Ra_c^{1.02}$.

It is of course plausible that fingers appear once the stabilizing temperature
gradient is strong enough and the density ratio exceeds some threshold value. It
is however very surprising that a stabilizing thermal buoyancy as small as 1/30
of the destabilizing chemical buoyancy ought to be enough to replace the
convection roll with fingers. We are therefore led to search for alternative
explanations and mechanisms which may govern the transition.

We will start with a few criteria and hypotheses gleaned from previous
publications. Ref. \onlinecite{Schmit11} has shown that in an infinitely
extended medium in which fingers are equivalent to elevator modes, the fastest
growing instability of the linearly stratified ground states is of the finger
type for $|\Lambda| > 0.9$ at the $Sc$ and $Pr$ of the experiment. This includes
cases with $|\Lambda|<1$, but does not explain why fingers are observed in the
experiment for $|\Lambda|$ as small as 1/30.

Ref. \onlinecite{Stern69} and ref. \onlinecite{Holyer1980} showed that fingers become unstable to long internal
waves if $Re_f \geq \sqrt{\frac{2Ra_Td^4}{3PrL^4}}$. This relation is obtained from equations
(2.1), (2.2), (2.3) and (4.4) of ref. \onlinecite{Stern69} and equation (1.1) of ref. \onlinecite{Holyer1980} using our notations.
In the experiment, the data never satisfy this
relation (see table \ref{table1}). We will thus disregard this criterion. Holyer
\cite{Holyer1984} extended the stability analysis to include short
wavelength disturbances. These disturbances can lead to faster growth rates of
the instability than long wavelength modes. However, this work did not result in an analytical
criterion and it applies to a basic state in the classical finger regime in
which the total density stratification is stable, so that this work is not
immediately useful here.

\begin{figure}
\includegraphics[width=8cm]{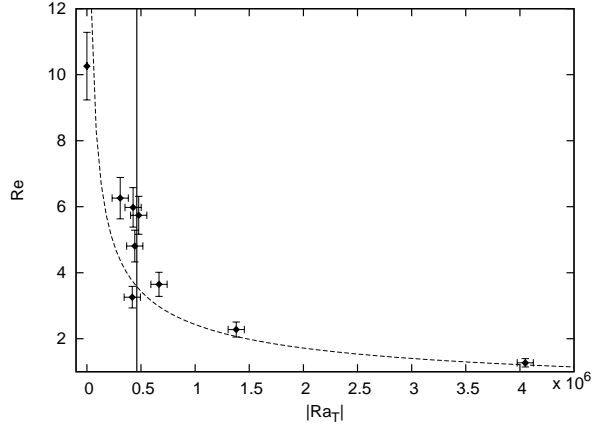}
\caption{$Re$ as a function of $|Ra_T|$ for $Ra_c=2.43 \times 10^{9}$. The
vertical line marks the transition and the dotted line is given by the relation
(\ref{eq_Re_Ra}).}
\label{fig_RevsRa}
\end{figure}

It was already noted in Ref. \onlinecite{Hage10} that the Sherwood number goes
to infinity according to eq. (\ref{eq_Sh_Ra}) if $Ra_T$ tends to zero. Since the
temperature gradient is stabilizing in the present configuration, it was
speculated that fingers disappear and eq. (\ref{eq_Sh_Ra}) loses validity at the
$Ra_T$ at which the predicted $Sh$ exceeds the value $Sh$ takes in the absence of a
stabilizing temperature gradient. This speculation is disproved by fig.
\ref{fig_ShvsRa}. Despite the adverse temperature gradient, finger convection
can transport more ions than a convection roll in an isothermal fluid. The
maximum of $Sh$ is reached at the transition. Fig. \ref{fig_RevsRa} demonstrates
that the velocity amplitude follows a more intuitive behavior: The smaller
$|Ra_T|$, the larger $Re$. This is of course expected because narrow fingers
(which exist at large $|Ra_T|$) increase friction, and because the stabilizing
gradient generally opposes motion. The reason why $Sh$ can nonetheless have a
maximum is the structure of the velocity field: In a convection roll, vertical
velocities which can transport ions from one electrode to the other are only
found at the perimeter of the roll, whereas in fingers, up- and downward
transport occurs almost everywhere. The most efficient transport thus occurs in
the fastest fingers, i.e. at the transition between fingers and convection
rolls.

Let us next check the hydrodynamic stability of the finger regime. The Reynolds
number of individual fingers, based on vertical velocity and finger thickness,
is below 1 at the transition, with one exception where it is 1.32 (see table
\ref{table1}). It seems excluded that hydrodynamic instability within individual
fingers destroys the fingers. The velocity field of the whole collection of
fingers is reasonably approximated by what is sometimes called the Kolmogorov
flow, defined as $\bm v = v_{y0} \sin (\pi x/d) \hat{\bm y}$, where $\hat{\bm
y}$ is the unit vector in $y-$direction. This flow is stable \cite{Green74} if
$v_{y0} d/\nu < \pi \sqrt{2}$, or $Re_y < \pi$ (since $Re_y$ is
defined with the rms rather than the maximum value of velocity). According to
table \ref{table1}, this criterion is always met at the transition. Finally, the
Rayleigh number of the chemical boundary layer (which has the thickness 
$L/(2 Sh)$ and across which there is a drop in concentration of $\Delta c/2$) is
given by $Ra_c/(16 Sh)^3$ and is also listed in table \ref{table1}. This Rayleigh
number is always less than 250 at the transition. In summary, there is no
obvious way to mechanically destabilize the fingers.

\begin{figure}
\includegraphics[width=8.4cm]{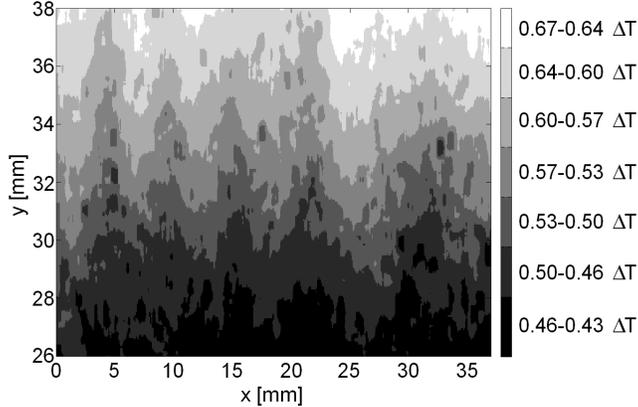}
\caption{Gray scale plot of the temperature field visualized in a vertical plane
with thermochromic
liquid crystals for $Ra_c=2.92\cdot10^{10}$ and $|Ra_T|=1.86\cdot10^7$.
Temperatures are grouped into six bins for clarity. The top and bottom plates
are located at $y=0\;$mm and $y=60\;$ mm, respectively.}
\label{fig_Isolinien}
\end{figure}

\begin{figure}
\includegraphics[width=8cm]{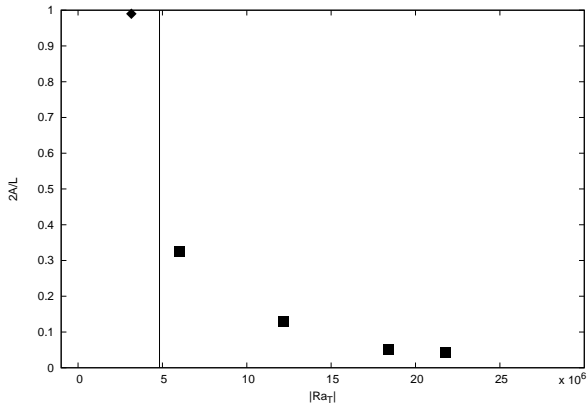}
\caption{The peak to peak amplitude $2A$ of the distortion of the isotherm with
temperature $T$ given by $(T_{\rm{bottom}}-T)/\Delta T = 0.51$, as a function of
$|Ra_T|$ for $Ra_c=2.92 \times 10^{10}$, determined from experimental
visualization (filled squares). The diamond denotes a point obtained from the 
experiment without a detectable temperature gradient. Therefore this value was
arbitrarily set to $2A/L=1$.}
\label{fig_t_Verformung}
\end{figure}

We now turn to the temperature field. Fingers only occur within a stable
temperature stratification, and one may hypothesize that as $|Ra_T|$ is
decreased, the vertical velocity within the fingers increases and destroys more
and more the temperature gradient until there is no reason left to form fingers.
In order to test this hypothesis, the temperature field was visualized
experimentally with thermochromic liquid crystals.
An example of a temperature field obtained by this technique is sown in fig.
\ref{fig_Isolinien}. For slow convective motion, one expects the isotherms to be
distorted by advection in a sinusoidal fashion as seen in this figure. Since the
top and bottom boundaries remain at constant temperature, such a deformation
implies that the temperaure gradient is increased near the top plate in the
updrafts and near the bottom plate in the downflows. The advection of the
temperature field therefore
does not globally reduce the stable stratification as long as the amplitude of
the distortion of the isotherms is less than the cell height. In a well mixed
state, the temperature outside the boundary layers adjacent to the top and
bottom plates is everywhere equal to the aritmetic mean of the temperatures of
the top and bottom plates, so that the amplitude of the deformation of the
isotherm with the temperature $T$ defined by 
$(T_{\rm{bottom}}-T)/\Delta T = 0.5$
must be larger than half the cell height. The difference $2A$ between the maximum and
minimum height of the isotherm with $(T_{\rm{bottom}}-T)/\Delta T = 0.51$ was
extracted from pictures like fig. \ref{fig_Isolinien} for different $Ra_T$ at
$Ra_c=2.92 \times 10^{10}$ and is shown in fig. \ref{fig_t_Verformung}.
As the transition is approached from within the finger regime by reducing
$|Ra_T|$, the distortion of the isotherms increases, but at the transition,
there is still a significant temperature gradient left, whereas there is
no detectable temperature stratification below the transition in the convection
roll.  This is indicated in fig. \ref{fig_t_Verformung} by an experimental point
wich we assign the value of $2A/L=1$. There is
therefore no evidence that a reduction of the stabilizing gradient leads to the
disappearance of the fingers, since when the transition occurs, there is still
a gradient present which in parts of the cell is even larger than the imposed
gradient.

Apart from setting up a stabilizing gradient, temperature has another criterion
to fulfill for fingers to form: Heat needs to diffuse fast enough. Fingers form
because heat is exchanged between neighboring fingers, but ions are not. Heat
therefore has to diffuse over a distance larger than the finger thickness during
the transit time from one electrode to the other. The ratio of the diffusion
length to the finger thickness is given by $\frac{1}{d}\sqrt{\kappa L/V}$ and is
also listed in table \ref{table1}. Fingers should disappear if 
$\frac{1}{d}\sqrt{\kappa L/V} \approx 1$. Inserting the scalings for $d$ and $V$
from eqs. (\ref{eq_d_Ra}) and (\ref{eq_Re_Ra}), this predicts the transition to
occur on a line in the $|Ra_T|,Ra_c-$plane defined by
$|Ra_T| = 6.6 \times 10^{-6} Pr^{6/7} Ra_c^{22/21}$, where the prefactor results
from the requirement that $\frac{1}{d}\sqrt{\kappa L/V}$ be exactly 1.
The exponent 22/21 is indistinguishable from the
one found in fig. \ref{fig_Phasenraum} because 22/21=1.05. With this exponent,
the best fitting prefactor deduced from the data in fig. \ref{fig_Phasenraum} is $6.9 \times 10^{-6}$.

While the exponent of $Ra_c$ is nearly the same as the one obtained from the
transition criterion $\Lambda=const.$, the two criteria have different
dependencies on $Pr$ and $Sc$, but these parameters cannot be varied
significantly in the present set-up. The ratio of heat diffusion length and
finger thickness varies by a factor of two along the transition line according
to table \ref{table1}, but these variations are not systematic and are
attributable to uncertainties in the measurements. 

We therefore arrive at the picture that fingers exist as long as the diffusion
of heat reaches across fingers. As the transitional $|Ra_T|$ is approached from
above, velocities increase, transit times decrease and fingers thicken, until a
point is reached where the thermal diffusion length during the transit time equals the
finger width and fingers disappear. This leaves the question why fingers appear in a top heavy fluid
once a convection roll has been established and $|Ra_T|$ is increased beyond the
transition at constant $Ra_c$. We are thus reminded of the fact that we lack a
definitive understanding for why convection rolls form in high Rayleigh number
convection. One could equally well imagine convection in the form of many
independent plumes crossing the cell much as bubbles in a boiling pot of water
without a large scale circulation. We know that high Rayleigh number convection
does result in large scale convection cells, but as the present experiments
show, a small disturbance of 1/30 th of the buoyancy force is enough to disrupt
the convection roll and to replace it with a small scale convection structure.


\begin{table}\centering
\begin{tabular}{|c|c|c|c|c|c|c|c|}
\hline
$Ra_c$ & $|Ra_T|$  & $\Lambda$ & $\ \ Re\ \ $ & $\ \ Re_f\ \ $ & $Ra_{c,\lambda}$ & 
$\sqrt{\kappa L /(v d^2)}$ & $\sqrt{\frac{2}{3}\frac{Ra_T}{Pr}\left(\frac{d}{L}\right)^4}$ \\ \hline
\hline
 $1.00\cdot10^{09}$ & $1.33\cdot10^{5}$ & 0.030 & 03.6 & 0.59 & 138 & 1.45 & 1.87\\
		    & $1.15\cdot10^{5}$ & 0.026 & 04.1 &      & 140 &      & \\
\hline
 $2.43\cdot10^{09}$ & $4.79\cdot10^{5}$ & 0.044 & 05.7 & 0.73 & 065 & 1.03 & 4.34\\
 		    & $4.42\cdot10^{5}$ & 0.041 & 04.8 &      & 085 &      & \\
\hline
 $1.73\cdot10^{10}$ & $1.78\cdot10^{6}$ & 0.024 & 04.2 & 0.39 & 112 & 1.66 & 4.52\\
 		    & $1.56\cdot10^{6}$ & 0.021 & 04.7 &      & 112 &      & \\
\hline
 $2.92\cdot10^{10}$ & $4.85\cdot10^{6}$ & 0.037 & 18.9 & 1.32 & 110 & 1.01 & 4.77\\
		    & $4.82\cdot10^{6}$ & 0.037 & 34.9 &      & 119 &      & \\
\hline
 $1.39\cdot10^{11}$ & $2.05\cdot10^{7}$ & 0.034 & 10.8 & 0.63 & 192 & 1.69 & 5.15\\
	  	    & $1.96\cdot10^{7}$ & 0.032 & 09.7 &      & 191 &      & \\
\hline
 $1.85\cdot10^{11}$ & $2.83\cdot10^{7}$ & 0.035 & 08.8 & 0.60 & 159 & 1.66 & 7.58\\
	  	    & $2.47\cdot10^{7}$ & 0.031 & 14.5 &      & 181 &      & \\
\hline
 $2.16\cdot10^{12}$ & $4.51\cdot10^{8}$ & 0.048 & 25.2 & 0.62 & 240 & 2.28 & 5.18\\
		    & $3.98\cdot10^{8}$ & 0.042 & 29.9 &      & 245 &      & \\
\hline
\end{tabular}
\caption{Parameters for the cell adjusted to be near, both above and below, the transition between finger
convection and convection rolls: $Ra_c$, $Ra_T$, $\Lambda$, and $Re$ as
defined in the text, together with the Reynolds number based on finger
thickness, $Re_f=Re_y d/L$, chemical boundary Rayleigh number $Ra_{c,\lambda}=Ra_c/(16Sh^3)$, 
the ratio of heat diffusion length to finger thickness, $\sqrt{\kappa L /(v
d^2)}$ and the Stern criterion (see Refs. \onlinecite{Stern69} and \onlinecite{Holyer1980}).
No value is given for $Re_f$
and in the last two columns for points below the transition where no fingers exist.}
\label{table1}
\end{table}


\section{Conclusion}

The transition from finger convection to convection rolls in double-diffusive
convection was studied in an electrodeposition cell for $Pr \approx 8.7$ and $Sc
\approx 1970$ with stabilizing temperature and destabilizing concentration
gradients. Fingers are observed for density ratios 
$|\Lambda|=\alpha |\Delta T|/(\beta \Delta c) < 1$, i.e. when the system is
denser at the top than at the bottom. Fingers are replaced by convection rolls
only for $|\Lambda| < 1/30$. At the transition, the ion transport is larger than
without an adverse temperature gradient. From various experimental observations,
the most
plausible mechanism limiting the existence of fingers is heat diffusion. If the
stabilizing temperature gradient is too small, heat has insufficient time to
diffuse between neighboring fingers during the transit from one boundary to the
other, so that the difference between the diffusivities of heat and ions cannot
have an effect any more. This happens for $|Ra_T| \propto Pr^{6/7} Ra_c^{22/21}$, which
cannot be distinguished experimentally from $|\Lambda| = 1/30$ because the
exponent 22/21 is too close to 1, and the dependencies on $Pr$ and $Sc$ cannot
be tested. It is left for future, possibly numerical, work to determine the $Pr$
and $Sc$ dependencies of the stability limit of fingers.

Another open issue concerns the role of the boundaries. Visualizations suggest
that fingers are born as Rayleigh-Taylor instabilities in the chemical boundary
layers. If the stabilizing temperature gradient is large enough, these grow into
fingers, whereas if it is too small, they merge into a
convection roll. Finger convection in the oceans frequently occurs in vertically
stacked horizontal layers with layers of finger convection separated by layers
mixed by ordinary convection. There are no solid walls limiting the fingers in
this system and it remains to be seen whether fingers still exist for
$|\Lambda|$ as small as 1/30 in this case.


%

\end{document}